# Computing exponentially faster:
# Implementing a nondeterministic universal Turing machine using DNA

Andrew Currin[1,2,†], Konstantin Korovin[3*], Maria Ababi[4], Katherine Roper[3], Douglas B. Kell[1,2], Philip J. Day[5], Ross D. King[3,4,*]

1. SYNBIOCHEM, Manchester Institute of Biotechnology, The University of Manchester, United Kingdom.

2. School of Chemistry, The University of Manchester, United Kingdom.

3. School of Computer Science, The University of Manchester, United Kingdom.

4. Manchester Institute of Biotechnology, The University of Manchester, United Kingdom.

5. Faculty of Medical and Health Sciences, The University of Manchester, United Kingdom.

†. Joint 1st authors

* Corresponding author

**Abstract**: The theory of computer science is based around Universal Turing Machines (UTMs): abstract machines able to execute all possible algorithms. Modern digital computers are physical embodiments of UTMs. The nondeterministic polynomial (NP) time complexity class of problems is the most significant in computer science, and an efficient (i.e. polynomial P) way to solve such problems would be of profound economic and social importance. By definition nondeterministic UTMs (NUTMs) solve NP complete problems in P time. However, NUTMs have previously been believed to be physically impossible to construct. Thue string rewriting systems are computationally equivalent to UTMs, and are naturally nondeterministic. Here we describe the physical design for a NUTM that implements a universal Thue system. The design exploits DNA's ability to replicate to execute an exponential number of computational paths in P time. Each Thue rewriting step is embodied in a DNA edit implemented using a novel combination of polymerase chain reactions and site-directed mutagenesis. We demonstrate that this design works using both computational modelling and *in vitro* molecular biology experimentation. The current design has limitations, such as restricted error-correction. However, it opens up the prospect of engineering NUTM based computers able to outperform all standard computers on important practical problems.

**One Sentence Summary:** We experimentally demonstrate a Nondeterministic Universal Turing Machine (NUTM), NUTMs have exponential speedup over conventional and quantum computers on NP complete problems.

# Introduction

Universal Turing machines (UTMs) form the theoretical foundation of computer science (*1-13*): the *Church-Turing thesis* states that UTMs exactly define the concept of an algorithm - effective calculability. UTMs also play a fundamental role in science: the *Church-Turing principle* states that they are sufficient to simulate perfectly all physically realisable systems (*5,6,8*).

UTMs are an abstract mathematical concept, but the language that describes them begs a physical interpretation. Digital electronic computers physically embody UTMs, but differ from them in that they have bounded memory, may only run for a bounded amount of time, make errors, etc. (*9*). This tension between physical and abstract machines is at the heart of computer science.

The theory of computability investigates which problems a UTM can solve using unbounded space and time (*1-4,11-12*). The related theory of computational complexity investigates how much time and space are needed to solve particular problem classes (*2,4,11-15*). The complexity of an algorithm is its asymptotic worst-case use of a resource (space, time) as a function of the size of its input. A major conclusion of complexity theory is the 'feasibility thesis': that a natural problem has an efficient algorithm if and only if it has a polynomial-time (P) algorithm (*2,11,12,15*) (Fig. 1a). A function f : I → I is in the class P if there is an algorithm computing f and positive constants A, k, such that for every n and every $|x| \leq n$ the algorithm computes f(x) in at most $An^k$ steps.

The most significant concept in complexity theory is the class of nondeterministic polynomial time (NP) problems. Informally this is the class of decision problems where the *solution* can be verified in P time, i.e. membership propositions have short efficiently verifiable proofs (*2,11,12,14-17*). More formally a decision problem C is in the class NP if there is a function $V_c$ ∈ P and a constant k such that:

- If $x \in$ C then $\exists y$ with $|y| \leq |x|^k$ and $V_c(x,y) = 1$
- If $x \notin$ C then $\forall y$ we have $V_c(x,y) = 0$

A sequence *y* which 'convinces' $V_c$ that x ∈ C is often called a 'witness' or 'certificate' (*17*). Integer factorisation is an NP problem, for given an integer in its decimal representation its prime factors can be *verified* in P time (quadratic - $O(n^2)$ using schoolbook long multiplication), but there is no known P time algorithm to *find* an integer's prime factors. Many of the most important practical computational problems belong to the NP class, e.g. graph isomorphism, Boolean satisfiability, travelling salesman, graph colouring, etc. NP complete problems are the most difficult in NP, and all NP problems can be reduced to them in P time (*2,4,11,12,14,15*). This means that if you can solve any type of NP complete problem in P time then you can solve all NP problems in P time.

The NP class is commonly believed to be a strict superset of P, i.e. P ≠ NP; as it would seem generally harder to find a solution to a problem than to verify a correct solution (Fig. 1b). However this has never been proved, and the P = NP question is the arguably the most important open problem in mathematics (*11,15,18*). The problem is also of immense practical importance, for if P = NP it would essentially solve mathematics and transform science/engineering, but also have devastating consequences for activities that depend on cryptography for security, such as the banking system, the Internet, etc. (*13,15,18*).

It is important to distinguish the mathematical problem of the truth or falsehood of the proposition 'P = NP', and the practical problem of solving NP problems in P time (*9*). A mathematical problem is constrained by a given set of axioms and proof methods, whereas all possible physical means may be used to solve a practical problem. In this paper we do not address the P = NP mathematical problem, but instead present the physical design for a computer that has an exponential speedup over conventional and quantum computers on NP complete problems.

**Design of a Nondeterministic Universal Turing Machine (NUTM)**

The state of a UTM is defined by a tuple of symbols (*1,3*). In a standard (deterministic) UTM computation is a 1:1 relation that takes an input state to an output state, with computing halting if an accepting state is reached. A nondeterministic UTM (NUTM) differs from a UTM in that from one input state there may be multiple output states, i.e. computing is a 1:n relation (*2,3*). A now old-fashioned, but insightful, way to define the NP class is through the use of nondeterministic UTMs (NUTMs): the NP class it is the set of problems that a NUTM can solve in P time: $NP = \bigcup_{k \in \mathbb{N}} NTIME(n^k)$ (*2*).

The most common interpretation of how a NUTM solves a NP problem in P time is through *serendipity* (*7,10,11*): in each state it correctly guesses which of the output states to choose so as to reach the accepting state most quickly. Clearly this interpretation precludes the possibility of a physical NUTM, and one reads that they are 'magical' (*7*), 'hypothetical' (*10*), 'fictitious' (*11*), etc. Our alternative *replicative* interpretation is that a NUTM is a UTM that can reproduce itself, and thereby follow all computational paths in parallel, with the computation ending when one path reaches the accepting state. Such a machine is physically implementable.

We use a Thue rewriting system to implement a NUTM. Thue systems are a model of computation with equivalent power to Turing Machines (*2,3,19-22*). Formally a Thue system is the presentation of a monoid (*19*). Informally a Thue system is a set of rules of the form *w* ↔ *u* where *w, u* are strings in a finite alphabet of symbols. A string e.g. *v w v'* can be rewritten by the rule above to give *v u v'*. The application of a Thue rule to a string therefore produces a new string – equivalent to change of state in a UTM (Fig. 2a). The starting-state (program) is a specific string, as is the accepting-state. The execution of a Thue program consists of repeated application of Thue rewrite rules until an accepting state is produced (Fig. 2b). It is possible to translate any Turing machine into a Thue system, and vice-versa (*3,19*). The order and position of application of Thue rules is naturally nondeterministic: multiple Thue rules may be applied to a string, and individual Thue rules may be applied to multiple positions in a string (Fig. 2b). We implement the Thue system shown in (Fig. 2a), which is universal, i.e. it has undecidable (word) problems (*6,19-23*).

We use DNA computing to implement a Thue NUTM (Fig. 3a). The goal of building a molecular scale UTM has been pursued for over fifty years (*24*), with the most common molecule suggested being DNA (*24-30*), but proteins has also been proposed (*31*). DNA is an excellent medium for information processing and storage (*29,30*). It is very stable, as the sequencing of ancient DNA demonstrates. It can also reliably be copied, and many genes have remained virtually unchanged for billions of years. The foundational work on DNA computing was that of Leonard Adleman (*25*). He demonstrated the solution of a seven-point Hamiltonian path (a NP-complete problem) by generating a set of random DNA sequences (possible solutions), and selecting a solution from the set. Our work is an advance on this and all other work on molecular computing as we present the first physically demonstrated molecular UTM design, which means that there is no need for hardware redesign for different problems.

**Implementation of a DNA NUTM**

In our NUTM starting-states (programs) and accepting states (read-outs) are sequences of DNA that encode strings of Thue symbols (Fig. 3b). The physical architecture of the computer, with a mixing chamber, and editing chambers for each for different Thue rule/direction, ensures that every Thue rule is applied to every NUTM state (Fig. 2c). To physically implement each Thue rewrite rules we use a novel combination of Polymerase Chain Reactions (PCR) to copy state, and site directed mutagenesis (SDM) to change state (*26*). This approach ensures that all possible applications of a Thue rule are made.

The mechanism of the NUTM depends on the specificity of molecular binding – as do living organisms. The Boltzmann distribution $F(state) \propto e^{\frac{-E}{kT}}$ determines the frequency of molecular states of differing energies (E): higher energy states are exponentially less likely to occur than lower energy ones. The energy of DNA binding depends on sequence similarity (*32,33*), so the probability of undesirable DNA bindings can be made exponentially less likely through sequence design - although this is constrained by upper-limits on the temperature of DNA duplexes, etc.

To write programs (initial states) we use DNA synthesis technology. Accepting-states are specific sequences of DNA that contain identifying certificates, and corresponding answers to the computation. We require that the accepting states be recognised from among a potential exponential number of other states generated by the NUTM (Fig. 2b). This is feasible thanks to the Boltzmann distribution of binding energies, and because PCRs enable an exponential amount of complementary sequences to be produced in linear time. In our development work we read out accepting states directly using DNA sequencing. Other techniques are applicable, for example labelled complementary sequence to first identify the certificate, then sequencing to determine the result of the computation.

It is helpful to divide the task of applying a single Thue rule/direction into two steps: recognition, and rewriting. This separates the more thermodynamically challenging step of recognition, from the technically more complex step of rewriting. In rule recognition all antecedent strings of a given Thue rule are identified from among arbitrary complex strings of Thue symbols, and marked with a 'clamp' sequence. This clamp sequence is designed to be distinct from any DNA sequence encoded by Thue symbols, and thereby provide the specificity of binding required for rewriting. To insert the clamp sequence we use DNA oligonucleotide primers: these have at their 3' terminus a recognition sequence (the reverse complement of the antecedent of the rewrite rule), and at their 5' end the clamp sequence. The PCR products from these primers encode the clamp sequence adjacent to the target symbol sequence. This type of insertion procedure is a well-established site-directed mutagenesis technique (SDM) (*34,35*).

We have established *in vitro* that this recognition procedure works reliably. We have shown that we can recognise specific symbol string combinations and insert clamp sequences adjacent to them (**ec**, **ce**, **ae**, **ba**) in a Thue program (DNA template) containing multiple symbol combinations (Fig. 4). For the cases of **ec**, **ce**, **ae**, as expected, only one molecular weight (MW) band was produced. Sequencing demonstrated that the correct rule antecedent strings were identified, i.e. with the clamp sequence inserted at their 5' ends. For the **ba** symbol string, which occurs twice in the Thue program, as expected, we detected two different MW bands, and sequencing revealed that both possible rule antecedent strings were correctly identified (Fig. 4). This demonstrates nondeterministic rule recognition.

It would have been prohibitively expensive and time-consuming to physically demonstrate recognition against a background of all possible mismatching strings. We therefore applied computational modelling to demonstrate the specificity of recognition. The Gibbs free energy (G) of the hybridisation of DNA sequences to each other can be modelled with good accuracy (*32*). To calculate these estimates we used the UNAFold software (*33*). For each rewrite rule plus clamp we computationally verified that the binding that perfectly matches the rule antecedent sequence is energetically favourable (lower $\Delta G$) compared to binding with any other possible string of Thue mismatching symbols (see Supplementary material). This modelling approach is conservative as it is not generally the case for a Thue program that all Thue symbol strings may be produced, and because PCR depends on 3' binding, so the contribution of the 5' clamp is relatively unimportant.

We use SDM to make the changes of state required to implement Thue rewriting rules. As it is difficult to directly implement the complex DNA editing required for the universal Thue rules, we decomposed the rules into basic operations that can be directly implemented (see Supplementary material). These basic operations can then be arranged in different ways ('microprograms') to execute arbitrary complex Thue rules, and hence a variety of representations of an NUTM. The microprograms use a combination of symmetric and asymmetric PCRs to support the repeated targeting of multiple positions simultaneously (*36,37*). In physical terms the basic operations are DNA hydbridisations, where the new sequence (encoded by a DNA primer) mismatches and binds with an existing template, with the products of the reaction encoding the new sequence.

All the microprograms follow a similar schema: a series of mismatching symmetric and asymmetric PCR operations that implement the designed DNA sequence changes. Each PCR operation generates a specific intermediate sequence that is either used as a template or megaprimer for subsequent operations. In all the microprograms the first operation is insertion of the clamp. The second operation is change of the spacer sequence from **s** to **s'**, which serves to further mark the position of rewriting and strengthen the binding of mismatching primers. (In our current *in vitro* implementation clamp insertion and spacer change are combined.) DNA edits (insertions/deletions/swaps) are first made using symmetric PCRs to generate double-stranded DNA products (using the corresponding **end** (reverse) primer) - the edits being made to the truncated clamped sequence. Asymmetric PCRs are then used to generate megaprimers (single-stranded DNA product) that retain the required sequence changes, but have the clamp removed. Finally the megaprimers are used to introduce the edits into the full-length DNA sequence - using the megaprimer and corresponding **start** (forward) primer.

There are three types of Thue rewriting edits: transpositions, insertions, and deletions (Fig. 2a). To demonstrate that our SDM method is sufficient to implement transpositions we used as examples the microprograms: **ce → ec** (Fig. 5a), and **ec → ce** *(see Supplementary material)*; for both microprograms we show the *in vitro* PCR steps, and the experimental evidence for the correct transformations. The universal Thue rules 1-4 require such transpositions (Fig. 2a).

To demonstrate insertions we used the microprogram **ec** → **eca** (Fig. 5b), and for deletions the microprogram **cea** → **ce** (Fig. 5c). The universal Thue rule 7 requires such insertions and deletions (Fig. 2a). Insertion/deletion edits require that the hybridised DNA 'loops', either in the template (for deletion), or the primer (for insertion). In all the microprograms the most difficult case occurs when there are repeats of the Thue symbol to be swapped/inserted/deleted, as the primer and template could hybridise in an incorrect conformation. To overcome this a non-coding symbol (**x** or **z**) is inserted first which is then changed to its new symbol combination.

The most complex universal Thue rules are 5 & 6, as these involve transpositions, insertions, and deletions (Fig. 2a). To demonstrate that this form of universal rule can be implemented using our methodology we used as an example rule 5: **ce** ↔ **eca** (*see Supplementary material*). This rule can be implemented by integrating and adapting the above microprograms: **ce** → **ec** → **eca**; **eca** → **cea** → **ce** into a singe workflow. Taken together these results demonstrate that all the Thue rules required for a NUTM can be physically implemented using DNA mutagenesis.

**Discussion**

DNA computing trades space for time: 'there's plenty of room at the bottom (*24*). This gives it potential advantages in speed, energy efficiency and information storage over electronics (*25,29,30,38*): the number of operations for a desktop DNA computer could plausibly be ~$10^{20}$s (~$10^3$ times faster than the fastest current super-computer); it could execute ~$2 \times 10^{19}$ operations per Joule (~$10^{10}$ more than current computers); and utilise memory with an information density of ~1 bit per $nm^3$ (~$10^{12}$ more dense than current memory). These advantages mean that it is feasible that a DNA NUTM based computer could outperform all standard computers on significant practical problems (*39*).

Our design for a NUTM physically embodies an abstract NUTM. Due to noise the correspondence between a NUTM and its DNA implementations is less good than that between UTMs and electronic computers. Although noise was a serious problem in the early days of electronic computers (*40*), it has now essentially been solved. Compared to Quantum Computers (QCs) noise is far less a problem for DNA NUTM (*29,30*) as there are multiple well established ways to control it, e.g. use of restriction enzymes/CRISPR to eliminate DNA sequences that do not encode symbols, use of error-correcting codes, repetition of computations, amplification of desired sequences, etc. Most significantly, when a NP problem is putatively solved the answer can be efficiently checked using an electronic computer in P time.

A major motivation for this work is to engineer a general-purpose way of controlling cells. The natural way cells are controlled is a very complex combination of DNA, RNA, protein, and small-molecule interactions (supplemented by epigenetics, etc.) with multilevel control implemented through specific chemical interactions. This makes cells very difficult to reprogram. Synthetic biology has sought to control cells through the design of simple switches, circuits, etc. and has some notable successes (e.g. (*41*)). However, we argue that a more radical and general approach is required: a DNA UTM. This would in principle enable arbitrary biological processes to be programmed and executed. The UTM could receive biological signals from the environment through interaction with transcription factors, etc. It could also use as effectors RNA/proteins generated using special sequences and RNA polymerase, etc. Our current *in vitro* implementation of a NUTM is not suitable for this. However, it would seem possible to implement the core ideas in a biological substrate. One way to do this would be to use plasmids as programs, and rolling circle replication.

The theory of computational complexity treats time and space as fundamentally different: space is reusable while time is not. The resource limitation in a physical NUTM is space, not time. The speed of the computation increases exponentially, while the amount of space is polynomially bound: the light-cone is cubic, and the holographic bound on the maximum amount of information in a volume of space is quadratic (*13,42*). Computation therefore resembles an explosion. Although DNA nucleotides are very small (Avogadro's number is ~6 x $10^{23}$) they are still of finite size, and this restricts the size of NP problem that a DNA NUTM could practically solve before running out of space - the Earth has ~$10^{49}$ atoms, and the observable Universe only ~$10^{80}$. We therefore argue that what protects cryptographic systems from being broken is not lack of time, as is generally argued (*11,13,43*), but lack of space.

Computation in a deterministic UTM is in principle reversible, i.e. there is no lower bound on the amount of energy required per operation (*44*). It is unclear whether NUTM computation is reversible in P time. This question is of importance in relation to power constraints in NUTMs, and to the P = NP question.

Given the prospect of engineering a NUTM it is natural to consider whether machines could be physically engineered for other complexity classes. A problem is a member of the class co-NP if and only if its complement is in the complexity class NP (Fig. 1b). The definition of NP uses an existential mode of computation: if any branch of the computation tree leads to an accepting state, then the whole computation accepts. The definition of co-NP uses a universal mode of computation: if all branches of the computation tree lead to an accepting state then the whole computation accepts. It would therefore be straightforward to adapt out NUTM design to compute co-NP problems: all accepting states are removed from the mixing vessel.

It would also be straightforward to add randomisation to a physical NUTM (through the use of thermal noise). The class BPP (Bounded-Error Probabilistic Polynomial-Time) is the class of decision problems where there exists a P time randomised algorithm (*13*). Although the relationship between BPP and NP is unknown, it would seem computationally useful to generate an exponential number of randomised UTMs in P time, for example for simulations.

The complexity class PSPACE consists of those problems that can be solved by a Turing machine (deterministic or nondeterministic) using a polynomial amount of space (Fig. 1b). It is a superset of NP, but it is not known if this relation is strict i.e. if NP = PSPACE. In an NUTM all the computation is in a sense local: forks with no communication between computational paths. We hypothesise that a requirement for local computation is a fundamental definition of the NP class. In contrast, a physical PSPACE computer would seem to require highly efficient communication between computational paths, which seems challenging. We therefore conjecture that it is physically impossible to build a computer that can efficiently solve PSPACE complete problems.

Most effort on non-standard computation has focussed on developing QCs (*5,13,45*). Steady progress is being made in theory and implementation, but no universal QC currently exists. Although abstract QCs have not been proven to out-perform classical UTMs, they are thought to be faster for certain problems (*5,13,45*). The best evidence for this is Shor's integer factoring algorithm, which is exponentially faster than the current best classical algorithm (*45*). While integer factoring is in NP, it is not thought to be NP complete (*11*), and it is generally believed that the class of problems solvable in P time by a QC (BQP) is not a superset of NP (*13,18*).

NUTMs and QCs both utilize exponential parallelism, but their advantages and disadvantages seem distinct. NUTM's utilize general parallelism, but this takes up physical space. In a QC the parallelism is restricted, but does not occupy physical space (at least in our Universe). In principle therefore it would seem to be possible to engineer a NUTM capable of utilizing an exponential number of QCs in P time.

Advocates of the Many-Worlds interpretation of quantum mechanics argue that QCs work through exploitation of the hypothesised parallel Universes (*9,13,46*). Intriguingly, if the Multiverse were a NUTM this would explain the profligacy of worlds.

**Acknowledgments**

R.D.K. would like to thank Steve Oliver for helpful discussions, and the ERC for the spur he received by their non-award of grant ERC-2013-AdG 339781. A.C. and D.B.K. thank the Biotechnology and Biological Sciences Research Council (BBSRC) for support (BB/M017702/1). KK would like to thank the Royal Society for their support provided by the URF Fellowship.

**Supplementary Materials**

1. Thue Rule Recognition and Implementation

2. Thue Rule Microprogramming Formalization.

3. DNA Binding Estimates.

Fig S1

**Figure Legends**

*Figure 1 Computational complexity.*

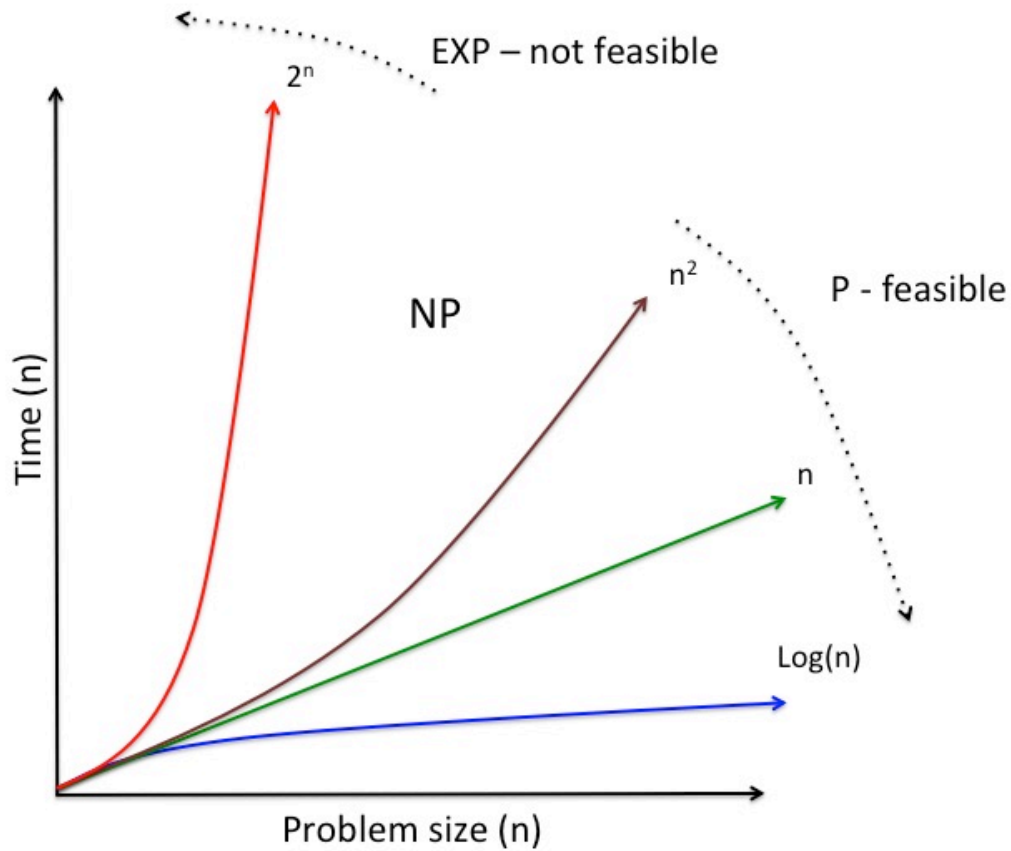

A) The feasibility thesis asserts that there is a fundamental qualitative difference between algorithms that run in Polynomial time (P time) (e.g. schoolbook multiplication), and algorithms that run in exponential time (EXP time) (e.g. position evaluation in a generalised game) (*2,11-15*). As problem size increases P time algorithms can still feasibly (efficiently) be executed on a physical computer, whereas EXP time algorithms cannot. The feasibility thesis also asserts that NP algorithms cannot feasibly be executed, but this is less clear as this assumes P ≠ NP.

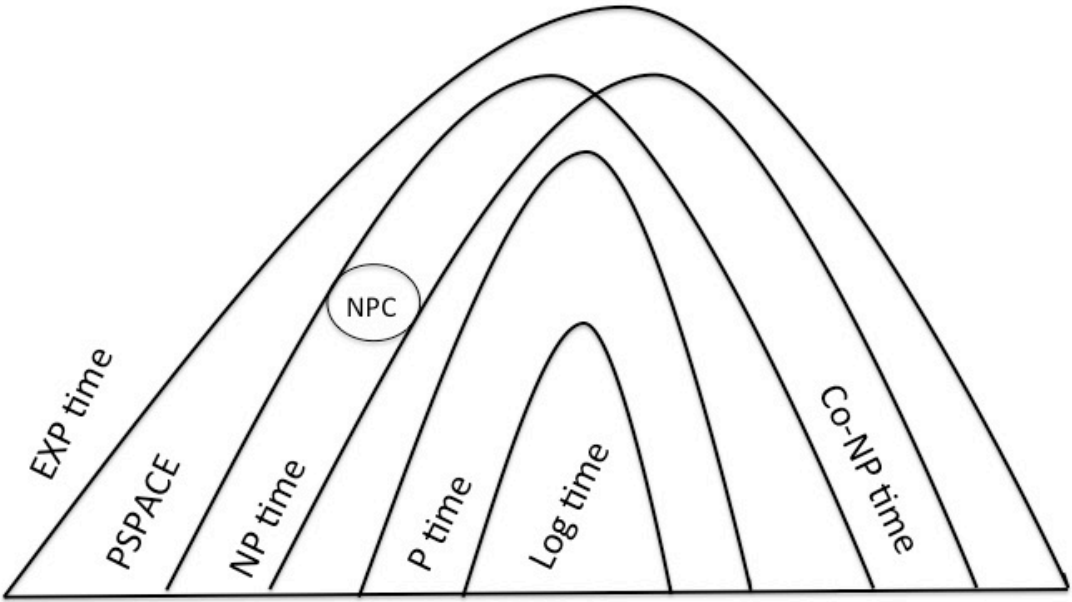

B) Complexity classes are related through the subset relationship: Log time ⊆ P time ⊆ NP ⊆ PSPACE ⊆ EXP time (*2,11-15*).  Little is known of the exact details of these relationships, e.g does P = NP?

*Figure 2. Thue systems.*

| | | | |
|---|---|---|---|
| ① | a c | ↔ | c a |
| ② | a d | ↔ | d a |
| ③ | b c | ↔ | c b |
| ④ | b d | ↔ | d b |
| ⑤ | c e | ↔ | e c a |
| ⑥ | d e | ↔ | e d b |
| ⑦ | c d c a | ↔ | c d c a e |

*A*) A set of universal Thue rules: rules 1-4 require symbol transposition; rule 7 requires symbol insertion (forward) and deletion (reverse); and rules 5 and 6 requires transposition, insertion (forward), and deletion (reverse).

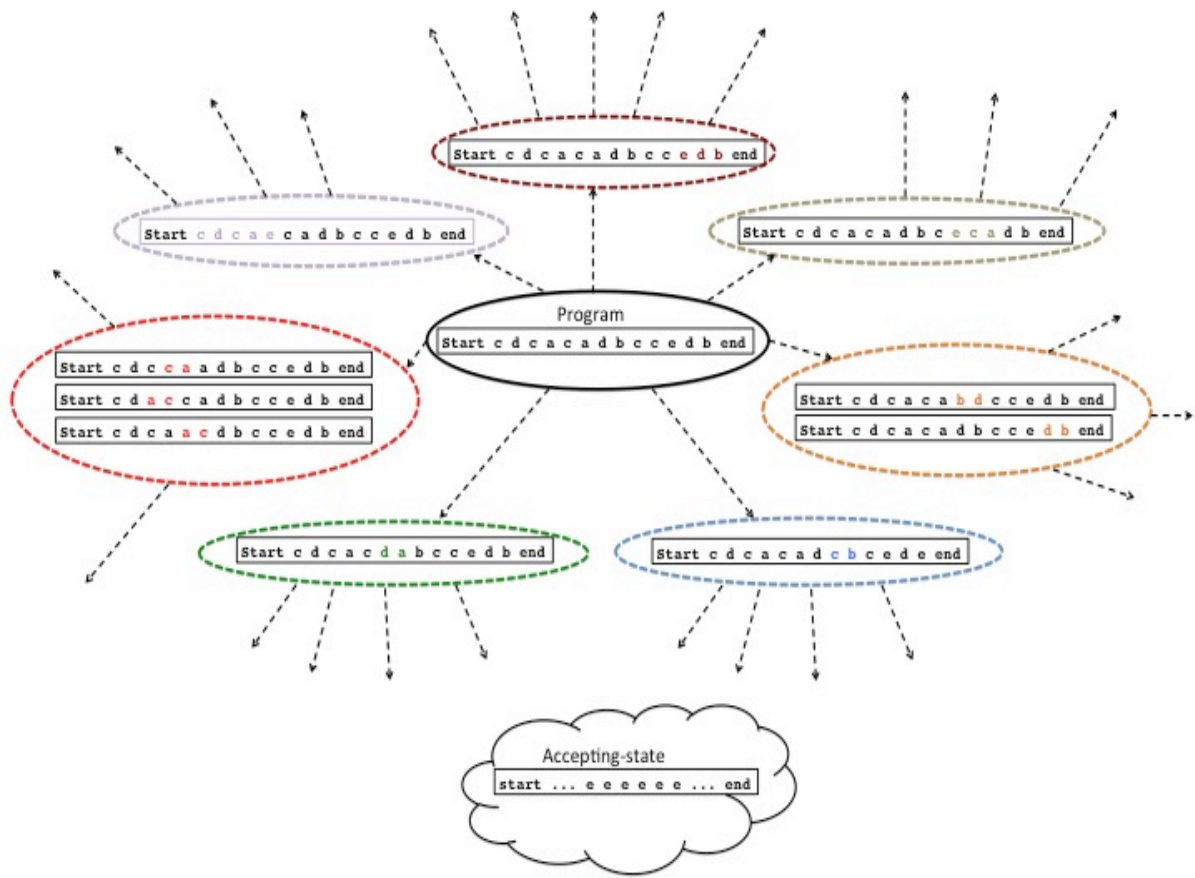

*B)* Example trace of the execution of a NUTM. The execution of NUTM is a tree of all possible computations. The root of the tree is the initial program. The child nodes of the root are the subsequent Thue sequences generated from the initial program by application of one of the seven Thue rules: note that the antecedent of a rule (e.g. **ca** – the reverse form of rule 1) may occur multiple times. Thue rules are recursively applied until the accepting state is produced, thus execution of a program generates a potentially exponential number of states in P time

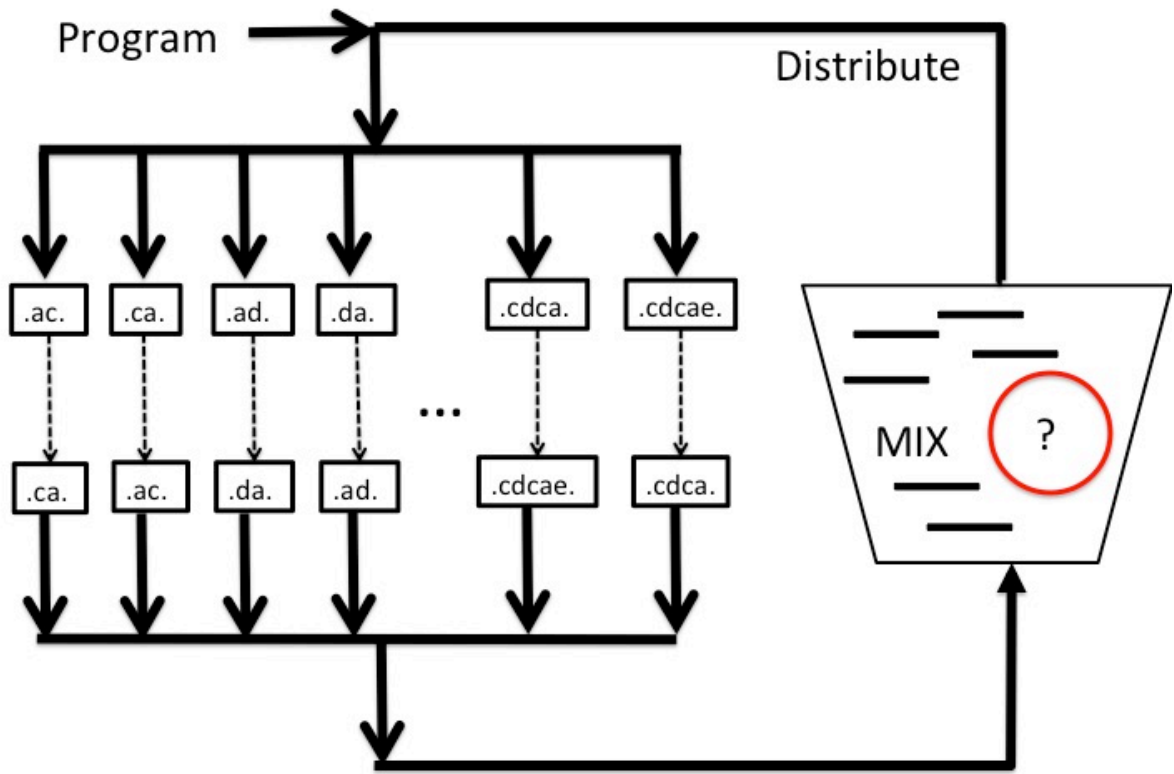

*C)* The physical NUTM consists of 14 paralleling executing processors (one each for both direction of the seven Thue rules) and a mixing vessel. Each processor executes a microprogram that implements a Thue rule in one direction. The resulting transformed DNA sequences are added back to the mixing vessel, mixed, and then distributed to the processors to continue the computation. In this way all possible combinations of computational steps are executed. The mixing vessel also contains a method of recognising accepting-states.

*Figure 3. DNA computing.*

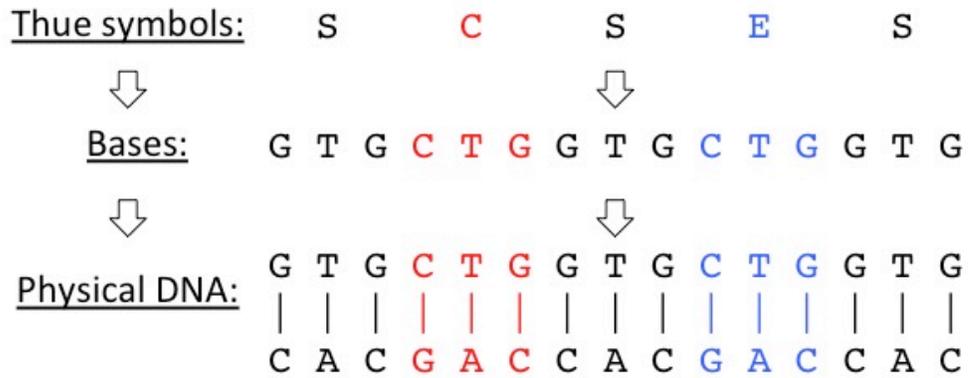

*A*) Three levels of symbol can be identified in our NUTM: physical DNA, bases, and Thue symbols.

| Category | Symbol | DNA sequence |
|---|---|---|
| Thue symbols | a | TCT |
|  | b | GCT |
|  | c | TGG |
|  | d | ACG |
|  | e | CTG |
| Spacers | s | GTG |
|  | s' | GCG |
| Clamp |  | GGAATGTCAACATGATA |
| Intermediate symbols | x | CGG |
|  | y | AAA |
| Delimiters | start | TCGAAGGTCG |
|  | end | TAAGGATCCGGCTGCTAAC |

*B*) We encode the five Thue system symbols **a**, **b**, **c**, **d**, **e** using triplets of DNA. This length was found to provide the best balance between symbol specificity and the ability to mismatch - intriguing the genetic code is also based on a triplet code (cypher). The use of triplets helps ensure that when performing PCRs with an annealing temperature in the range of 50-60°C, only the desired target sequences are amplified. A spacer symbol **s** (or **s`**) occurs between each Thue symbols to help enforce the specificity of desired mismatches. We require the marker symbol **clamp** for rule recognition. We use the intermediate symbols **x** and **y** to help ensure that unwanted cross-hybridisation of symbols do not occur. Finally, the symbols **start** and **end** delimit the program. Physically these delimiters are used as recognition sites primers for PCRs.

*Figure 4. Thue rule recognition.*

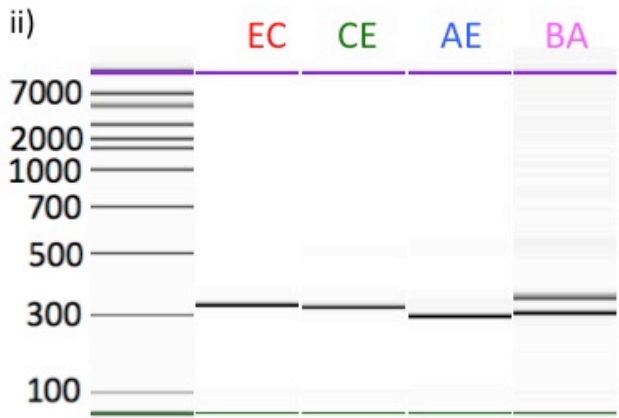

i) Sequence design of a DNA template encoding a string of 10 symbols separated by spacers. The symbol combinations **ec** (red), **ce** (green) and **ae** (blue) occurs once only within the string, whereas **ba** (pink) occurs twice. The complementary DNA primers consist of a clamp sequence followed by the symbol combination and flanking spacers. ii) Capillary electrophoresis analysis of PCR products for Thue rule recognition. Using the string template DNA described in (i), PCR reactions were carried out to insert a clamp sequence prior to each Thue rule symbol combination. For primers targeting the symbols **ec**, **ce** and **ae**, only one PCR product is created, while for **ba**, two products occur.

*Figure 5. Thue rule implementation*

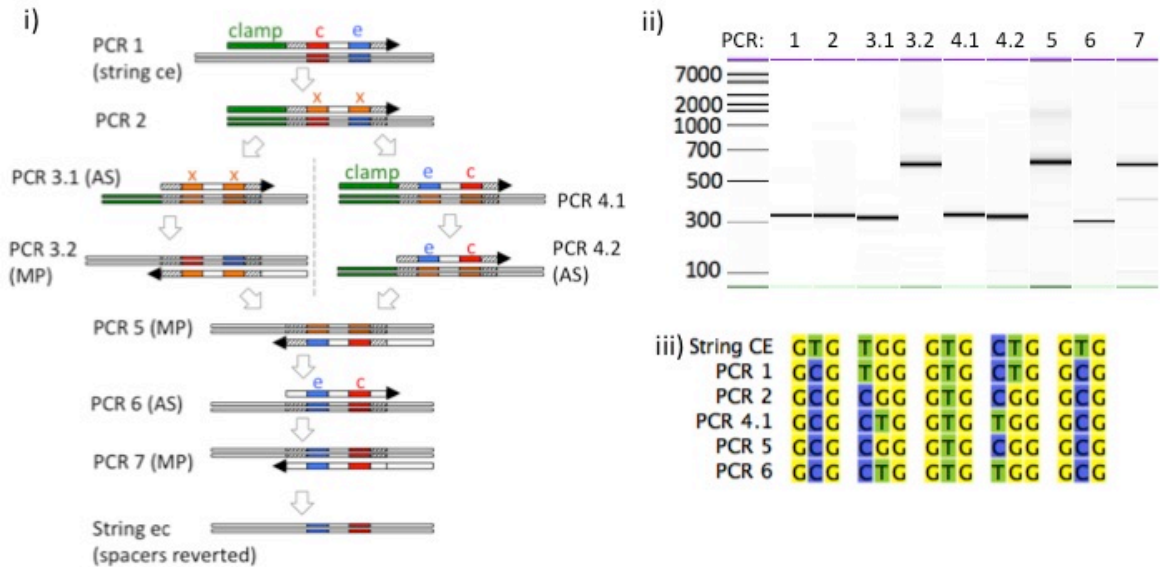

*A*) Microprogram for swapping **ce → ec**. i) The clamp sequence is first inserted and the outer **s** symbols changed to **s`** (PCR1). This clamp and spacer sequence is then bound by the primer in PCR2, which replaces the **ce** symbols with an **xx** sequence. Asymmetric PCR (AS) is then used to remove the clamp and create a megaprimer, which is used to replace **ce** with **xx** in the string (megaprimer PCR annotated "MP", PCR 3.1 and 3.2). In parallel the **xx** symbols are changed to **ec** by symmetric PCR (PCR 4.1), and then a megaprimer produced removing the clamp (PCR 4.2). This megaprimer is then employed to replace the sequence **xx** with **ec** in the string (PCR 5). Finally, the **s`** symbols are returned to **s** to complete the microprogram (PCR 6 and 7). ii) Capillary electrophoresis analysis of the PCR products from each PCR step. iii) Sequence alignment of DNA sequencing of the key steps in the microprogram.

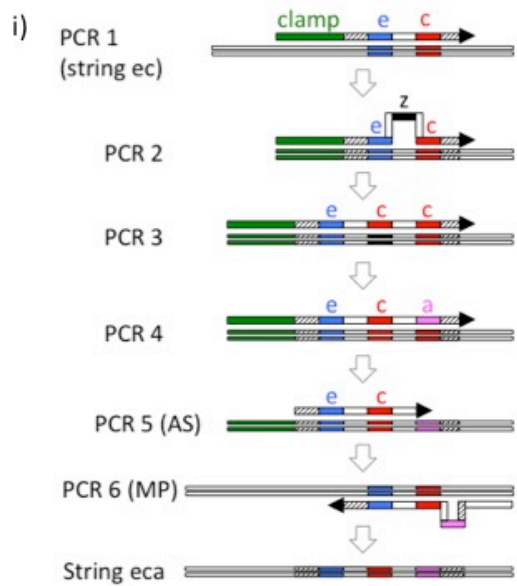
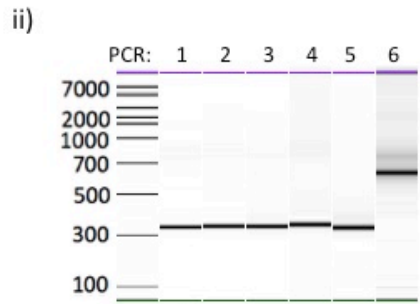

*B*) Microprogram for the insertion **ec** → **eca**. i) Following the recognition of **ec** the clamp is inserted and outer spacers changed (**s** to **s`**, PCR 1). A non-coding symbol **z** is then inserted, this exploits the strong binding for the existing **e** and **c** with the modified **s`** spacers (PCR 2), which promotes a loop to occur during DNA hybridisation. This symbol is then edited to **c** (PCR 3), and the other **c** changed to **a** (PCR 4). Asymmetric PCR is then used to generate a megaprimer for the **eca** sequence (PCR 5), which is then used to insert this new sequence into the tape (PCR 6). ii) Capillary electrophoresis analysis of the PCR products from each PCR step. iii) Sequence alignment of DNA sequencing of the key steps in the microprogram.

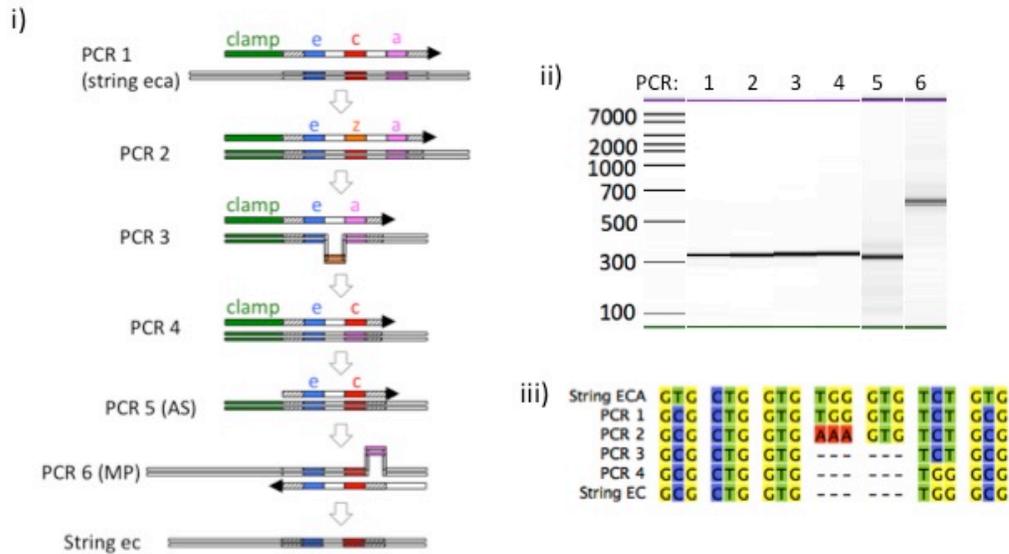

C) Microprogram for the deletion **eca** → **ec**. i) The clamp and altered spacers **s`** are inserted upon recognition of **eca** in PCR 1. The middle symbol (**c**) is edited to a non-coding **z** (PCR 2) before deletion of this symbol by recognition of **ea** in the subsequent step (PCR 3). As with the insertion microprogram, the strength of DNA hybridisation between the clamp, **s`e** and **as`** promotes the PCR primer to loop over the **z** symbol to delete it. Following deletion the symbol sequence **ec** is created (PCR 4), and the clamp removed by asymmetric PCR (PCR 5). Finally, the megaprimer is used to delete the original symbol **a** from the string. ii) Capillary electrophoresis analysis of the PCR products from the deletion microprogram. iii) Alignment of sequencing data from the key PCR steps.